\documentclass[english,aip,manuscript=article]{revtex4-1}
\usepackage{graphicx}
\usepackage{bm}
\usepackage{amsmath}
\usepackage[utf8]{inputenc}
\usepackage[T1]{fontenc}
\usepackage{mathptmx}
\usepackage{physics}
\usepackage{float}
\usepackage{hyperref}
\usepackage{xcolor}
\usepackage{soul}
\usepackage{comment}
\usepackage{array}
\makeatletter
\newcounter{subsubsubsection}[subsubsection]

\newcommand\subsubsubsection{\@startsection{subsubsubsection}{4}{\z@}%
  {-3.25ex\@plus -1ex \@minus -.2ex}%
  {1.5ex \@plus .2ex}%
  {\normalfont\normalsize\itshape}}
\newcommand\l@subsubsubsection{\@dottedtocline{4}{7em}{4em}}

\newcommand{\bR}{{\bm R}}

\newcommand{\bP}{{\bm P}}

\date{January 2026}
\begin{document}

\title{Electron Transfer, Diabatic Couplings and Vibronic Energy Gaps in a Phase Space Electronic Structure Framework}
\author{Zain Zaidi}
\email{zz4271@princeton.edu}
\affiliation{Department of Chemistry, Princeton University, Princeton, New Jersey 08540, United States}
\author{Xuezhi Bian}
\email{xzbian@princeton.edu}
\affiliation{Department of Chemistry, Princeton University, Princeton, New Jersey 08540, United States}
\author{Joseph E. Subotnik}
\email{subotnik@princeton.edu}
\affiliation{Department of Chemistry, Princeton University, Princeton, New Jersey 08540, United States}

\begin{abstract}
   We investigate the well-known Shin-Metiu model for an electronic crossing, using both a standard Born-Huang (BH) framework and a novel phase space (PS) electronic Hamiltonian framework. We show that as long as we are not in the strongly nonadiabatic region, a phase space framework can obtain a relative error in vibrational energy gap {and other vibronic matrix elements that} are consistently one order of magnitude smaller than what is found within a BH framework. In line with recent results showing that dynamics on one phase space surface can outperform dynamics on one Born-Oppenheimer surface, our results indicate that the same advantages should largely hold for curve crossings and dynamics on two or a handful of electronic surfaces, from which several implications can be surmised as far as the possibility of spin-dependent electron transfer dynamics.
\end{abstract}
\maketitle

\section{Introduction: Born-Huang Approach to Electron Transfer}

The standard theory of electron transfer (ET) is well known\cite{nitzanbook}. On the basis of a separation of time scales between nuclei and electrons, we assume that only two electronic states are important, a donor $\ket{D}$ and acceptor $\ket{A}$. These electronic states are further assumed to form a 
diabatic subspace that is immutable with respect to nuclear position. If we further assume the diabatic coupling is a constant (the Condon approximation), the resulting spin-boson model\cite{legget-spin-boson} can be manipulated so as to derive an electron transfer rate within a Marcus theory framework\cite{marcus_theory_1956} of the form:
\begin{equation}\label{eq:marcus-rate}
    k_{ET} = \frac{2\pi}{\hbar} \frac{|V_{DA}|^2}{\sqrt{4\pi E_R k_B T}} \exp\left( -\frac{(E_R+\Delta G^0)^2}{4 E_R k_B T} \right)
\end{equation}
In Eq. \ref{eq:marcus-rate}, the key parameters are the reorganization energy $E_R$, the diabatic coupling $V_{DA}$, and the energy difference between diabatic minima $\Delta G^0$.
\begin{figure}[h]
    \centering
    \includegraphics[width=0.85\linewidth]{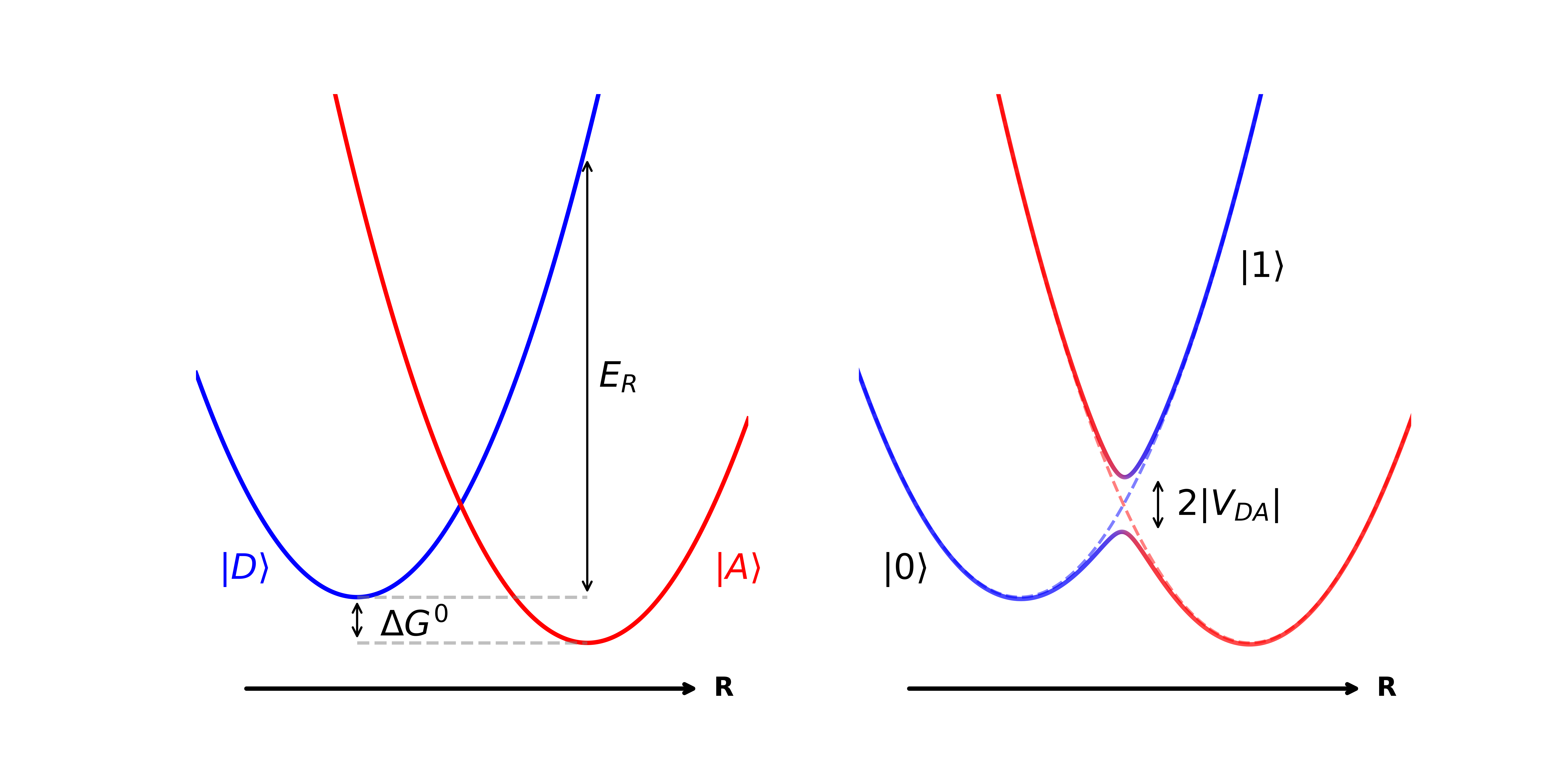}
    \caption{Diabatic (Left) and adiabatic (Right) Marcus parabolas with relevant quantities for nonadiabatic electron transfer. For the adiabats, diabatic character is labeled by color.}
    \label{fig:Marcus-Parabolas}
\end{figure}

Of course, all of the theory above is nothing more than a model. If one seeks to model electron transfer for a realistic system\cite{newtonmarshal_rev_1984}, one must necessarily run electronic structure calculations to extract the relevant electronic states and the corresponding parameters listed above, which inevitably leads to further considerations: 
(i) The relevant donor and acceptor states do change.
(ii) The entire subspace $\left\{ \ket{D},\ket{A} \right\}$ also changes, and the change in this subspace is quantified by the nonabelian Berry curvature\cite{mead_rev_nonabelian_1992}.
(iii) Intruder states can necessarily appear, which strongly breaks the notion of a two-state model.

Given the nuances described above, there are several ways to assess the accuracy of the spin-boson model that leads to Eq. \ref{eq:marcus-rate} vs. dynamics with a fully {\em ab initio} approach. On the one hand, one can simply compare calculated versus rates of electron transfer to experimental rates; often Marcus theory works quite well or well enough.\cite{closs_distance_1986,miller_intramolecular_1984,ohta-morokuma_marcus1_1986,koga-morokuma_MO-marcus_1993,beratan_rev-marcus_2019,liu-beratan_QD-marcus_2017,migliore-beratan_pcet-marcus_2014, onuchic-beratan_ET-marcus_1986,subotnik:2010:closs}
On the other hand, from a purely computational point of view, another means to check the accuracy of a model is to calculate and compare vibronic eigenvalues from the two different approaches (which can in principles be measured spectroscopically). This approach will be the focus of the present article. Below, our first goal will be to assess how well one can recover the lowest vibronic excitation energy of a system undergoing ET for a quantum system that is composed of two or three states -- as compared with an exact solution that arises from a formally infinite number of electronic states in principle.
Perhaps not so surprisingly, we will show that, if the mass difference between the nucleus and the electron is reduced further and further, larger and larger errors appear.

With this intuition in mind, our second goal below is to ask a very simple question: {for smaller mass differences, is the traditional framework for electronic structure, known as the Born-Huang\cite{born-huang_dynamical_1955} (BH) framework, the optimal means to generate a reduced subspace of electronic states?} In other words, is there perhaps a better two (or three) state basis of electronic states for understanding electron transfer? For this problem, 
{our intuition is that, by reducing the mass difference between electrons and nuclei, we will be able to address new conceptual challenges that arise for chemical systems with electronic degeneracies or near-degeneracies. After all, mathematically, the effect of a large mass difference is normally to create a large energetic gap between electronic states (which is the essence of the Born-Oppenheimer (BO) approximation); thus, vice versa, it should also hold that we can study small electronic gaps by reducing the mass ratio.}
Moreover, we have been strongly motivated by the our recent excursions into phase space (PS) electronic structure theory\cite{bian_phase-space-way_2025}.  Recently, when studying the Borgis model for hydrogen bonding and proton transfer, our research group has shown\cite{bian_PS-vibration_2025} that, if we parameterize the fast quantum states by both the position $\bR$ and momentum $\bP$ of the slow nuclear coordinates, and then extract vibrational energies with a Wigner transform\cite{case:2008:wigner_review}, single state PS energies can vastly outperform single state BH (BO) energies. Thus, the second premise of this paper is to explore whether the same advantages can be found when studying ET problems with more than one adiabatic surface of interest. Indeed, we will show below that, provided we are not in the extremely nonadiabiatic limit, one can generate improved vibronic energies using ``adiabatic'' and ``diabatic''  phase space 
electronic states.  Moreover, given the fact that single surface phase space dynamics conserve the total linear and angular momentum of a nuclear-electronic system\cite{tao_basis-free_nodate} (unlike Born-Oppenheimer dynamics\cite{bian_phase-space-rev_2026}), the results below suggest that when studying electron transfer, there should be strong advantages to using a PS (rather than the BH) framework.

An outline of the paper below is as follows. In Section \ref{Sec:II}, we review the theoretical framework for computing vibronic energies in the Born-Huang framework, using either the adiabatic or a diabatic basis; we further introduce an alternative phase space approach to study such strongly coupled vibronic problems. In Section \ref{Sec:III}, we review the Shin-Metiu model\cite{shin_multiple_1996,shin_nonadiabatic_1995} for quantifying such effects and we present our results,  which highlight the many strengths and the few weakeness of a PS approach. In Section \ref{Sec:IV}, we interpret the results of the previous sections, demonstrating that when PS succceeds, that successs is not coincidental but rather results from mixing many higher order states together; when PS fails, the problem must lie with the very nonlocal nature of a quantum transition in the nonadiabatic electron-transfer regime. Finally in Section \ref{Sec:V}, we conclude and point out future directions. 
As far as notation is concerned, operators in Hilbert space will be written with a hat, $\hat{O},\hat{R},\hat{P}$ -- hats are applied to both electronic and nuclear operators. Notably, electronic operators parameterized in Wigner space (i.e. partially-Wignerized nuclear-electronic operators) will be subscripted,  as  in $\hat{O}_W$.
Lower case letters index electronic states; $A$ is an  index for nuclei.

\section{Different approaches to Vibronic Energies}\label{Sec:II}

Consider a system of nuclei and electrons interacting, whereby an electron can be stabilized in two (or three) different configurations. For such nonadiabatic problems, 
we aim to solve a {vibronic} Hamiltonian of the following form 
\begin{equation}
    \hat{H} = \frac{\hat{P}^2}{2M} + \frac{\hat{p}^2}{2m} + \hat V(R,r) 
\end{equation}
where $\hat P$ represents nuclear momentum, $\hat p$ represents electronic momentum, and $V(R,r)$ is the interaction potential between electrons and nuclei. Let us now discuss  two canonically different frameworks for extracting vibronic energies.

\subsection{Born-Oppenheimer and Born-Huang Theory}

\subsubsection{Adiabatic Framework}

{In the Born-Huang framework, the initial step towards solving the vibronic problem is to consider the simpler electronic Hamiltonian which is parameterized by nuclear position $R$:
\begin{equation}\label{eq:adiab}
    \hat{H}_{el}(R) \equiv \frac{\hat{p}^2}{2m} +\hat V(r;R)
\end{equation}
A unitary operator $\hat{U}_{el}$ is defined such that it diagonalizes $\hat{H}_{el}$ at each nuclear position $R$.}
\begin{equation}
    \hat{H}_{el}(R) = \hat{U}_{el}(R) \hat{\Lambda}(R) \hat{U}_{el}^{\dagger}(R)
\end{equation}
{Here, $\Lambda$ is a diagonal matrix whose elements indicate the electronic energy for each nuclear position $R$. The eigenvectors that are columns of $\hat{U}_{el}$ are denoted adiabatic states. } We then expand the total molecular wavefunction in this basis of adiabatic electronic states, which is known as the Born-Huang (BH) expansion\cite{born-huang_dynamical_1955}. 
\begin{equation}
    \Psi(r,R) = \sum_j \Omega_j(R)\Phi_j(r;R)
\end{equation}
Here, $\Phi_j$ is the j$^{th}$ adiabatic electronic state and $\Omega_j(R)$ is the vibrational wavepacket on that state. 
{In this adiabatic basis, the total vibronic Hamiltonian can be written as:
\begin{equation}\label{eq:BH_vibronic}
    (\hat{H})_{ij} = \sum_k(\hat{P}\delta_{ik} - i\hbar \hat d_{ik}) \frac{1}{2M} (\hat P\delta_{kj}-i\hbar \hat d_{kj}) + \hat \Lambda_{ij}\delta_{ij}
\end{equation}
Here $\hat{d}_{ij} = d_{ij}(\hat{R})$ is the nonadiabatic coupling, sometimes referred in the literature as the derivative coupling\cite{cederbaum:review:conicalbook}: 
\begin{equation}
    d_{ij}(R) \equiv \bra{\Phi_i (R)}\frac{\partial}{\partial R}\ket{\Phi_j (R)}
\end{equation}
If we multiply out the quantities in parenthesis in  Eq. \ref{eq:BH_vibronic}, an equivalent expression is:
\begin{equation}\label{eq:BH_vibronic_expanded}
    (\hat H)_{ij} = \delta_{ij}\frac{\hat P^2}{2M} + \Lambda_{ij}\delta_{ij} - i\
    \hbar\frac{\hat P \cdot \hat d_{ij} + \hat d_{ij} \cdot \hat P}{2M} - \hbar^2 \sum_k \frac{\hat{d}_{ik}\cdot \hat d_{kj}}{2M}
\end{equation}
This form allows for a clear interpretation of the electronic energies $\Lambda_{ii}$ as effective potential energy surfaces for the nuclei while the final two terms in Eq. \ref{eq:BH_vibronic_expanded} represent the coupling between vibrational wavepackets $\Omega_i$ on different electronic surfaces.}

{Eq. \ref{eq:BH_vibronic_expanded} above is an exact representation of the Hamiltonian only if one includes a complete (and infinite) number of electronic states. Often, one is interested in the low-energy vibronic states of a system. In such a case, the derivative couplings between the ground adiabatic electronic state and other states $\hat{d}_{0j}$ are very small. The Born-Oppenheimer (BO) approximation is when one chooses to neglect the derivative coupling terms (the third and fourth terms) in Eq. \ref{eq:BH_vibronic_expanded}, so that lowest vibrational wavefunctions can be evaluated from the much simpler Schrodinger equation:}
\begin{equation}\label{eq:BO_vibronic}
    \left(\frac{\hat{P}^2}{2M} + \Lambda_0(R)\right)\Omega_n(R) = \varepsilon_n\Omega_n(R)
\end{equation}
Here $\Lambda_0(R)$ denotes the ground potential energy surface for the nuclei and $n$ indexes the vibrational wavefunctions.

For weakly coupled adiabatic surfaces, the BO approximation is an excellent starting point. 
However, when multiple surfaces are strongly coupled, {that is, when the derivative couplings are large and $|P\cdot d_{12}/M|$ is commensurate with or larger than the energy spacing $\Delta E = \epsilon_2-\epsilon_1$,
} a single electronic surface obviously cannot accurately capture the correct low-energy physics. {Moreover, because of the rapid change in the electronic character between two or more adiabatic states, the derivative couplings can be sharply spiked near avoided crossings, leading to numerical instabilities--as is often the case for electron transfer and proton-coupled electron transfer\cite{sharon_nonadiabatic-pcet_2005,sharon-PCET_rev-2010}.
Thus, working in the adiabatic basis is often unstable in practice.} 
{
Note that in the above analysis, we have considered the case of a single moving nucleus (as relevant for our application below in Sec. \ref{Sec:III}). In reality, the derivative couplings form a three-tensor for a polyatomic system, $d_{ij}^{A \alpha} \equiv \bra{\Phi_i}\frac{\partial}{\partial R_{A \alpha}}\ket{\Phi_j}$, for atom $A$ in the $\alpha$ direction, which can lead to many further complications as well.}

\subsubsection{Diabatic Framework}
As an alternative to the adiabatic BH framework, one often rotates such adiabatic states through an adiabatic-to-diabatic transformation (ADT) in order to promote a globally consistent electronic character. 
{In order to apply such a diabatic framework, one assumes that one need not include the entire (complete) basis of adiabatic states; for instance, one might assume that one can run calculations within a subset of the lowest 15 electronic states (and ignore all higher electronic states). }
Let us denote this privileged adiabatic subspace of electronic states as $S$. Applying an ADT is equivalent to generating a new basis of ``diabatic states'' by applying a unitary transformation $U$ on the electronic states at each point in nuclear space: 
\begin{equation}\label{eq:diab_states}
    \eta_i(r;R) = \sum_{k\in S} \Phi_k(r;R) \left(U(R)\right)_{ki}
\end{equation}
In this new diabatic basis, the electronic energy is no longer diagonal:
{\begin{equation}\label{eq:diab_rotation}
     \tilde{V}_{ij}(R) = \sum_{k,l\in S}\big{(}\hat U^{\dagger}(R)\big)_{ik}\left(\hat{\Lambda}(R)\right)_{kl}\left(\hat{U}(R)\right)_{lj}
\end{equation}}
{The derivative couplings are also transformed:
\begin{equation}\label{eq:diab_NAC}
    \hat{\tilde{d}}_{ij} = \sum_{k,l \in S} \left((\hat{U}^\dagger(R))_{ik}(\hat{d}(R))_{kl} (\hat{U}(R))_{lj}\right) + \sum_{k\in S} \left((\hat{U}^\dagger(R))_{ik} \left(\frac{\partial \hat{U}(R)}{\partial R}\right)_{kj} \right)
\end{equation}
}
{
Above and below, we use a tilde ($\sim$) to denote matrix elements in a diabatic representation.  The total Hamiltonian matrix elements in the diabatic basis are:}
{\begin{equation}\label{eq:H_exact_diab}
    (\hat{H})_{ij} = \delta_{ij}\frac{\hat P^2}{2M} + \hat{\tilde{V}}_{ij} - i\hbar\frac{\hat P \cdot \hat{\tilde{d}}_{ij} + \hat{\tilde{d}}_{ij} \cdot \hat P}{2M} - \hbar^2 \sum_k \frac{\hat{\tilde{d}}_{ik}\cdot \hat{\tilde{d}}_{kj}}{2M},\quad i,j\in S
\end{equation}
}
{Because Eq. \ref{eq:H_exact_diab} is related to Eq. \ref{eq:BH_vibronic_expanded} by the (unitary) ADT in the subspace $S$, they represent the same Hamiltonian. The key difference is that, in the diabatic Born-Huang representation, $\hat{\tilde{V}}$ is no longer a diagonal electronic matrix, and $\hat{\tilde{d}}_{ij}$ represents the nonadiabatic coupling in the diabatic basis as described in Eq \ref{eq:diab_NAC}. Ideally, one would hope that the transformation in Eq. \ref{eq:diab_states} would eliminate such nonadiabatic couplings in Eq. \ref{eq:H_exact_diab}. Indeed, it has been shown by Baer\cite{baer_adiabatic-to-diabatic_1975} that a criterion must be satisfied for the existence of such an ADT, and Mead and Truhlar\cite{mead-truhlar_diabatic_1982} further showed that this criterion cannot be satisfied in general. For a comprehensive study of diabatic states, see Refs. \citenum{littlejohn_parallel-transport_2022,shu_truhlar-diabatic_states-2022}.} 

Given this state of affairs, the past 40 years have seen the development of a variety of quasi-diabatic approaches that smoothly minimize the nonadiabatic couplings such that they can be neglected in calculations for most regions of space\cite{yarkony_derivative-couplings_1996,subotnik_NAC-diabatic-accounts_2015,shu_truhlar-diabatic_states-2022}. In our application below to the Shin-Metiu model, we will use  Boys/GMH localization diabatization \cite{boys1966quantum,edminston-ruedenberg_localized_1963,cave_GMH_1996,cave_GMH_1997, subotnik_boys_2008} whereby we maximize the distance between diabatic charge centers; this approach is particularly useful for charge-transfer processes. For electrons in 1D and a subspace of $K$ adiabatic states (i.e. dim($S$)=$K$), Boys/GMH is equivalent to finding the rotation matrix $U$ (in Eq. \ref{eq:diab_rotation}) that enforces:
\begin{equation}\label{eq:boys-condition}
    \bra{\eta_i}\hat r\ket{\eta_j} = \delta_{ij}, \quad \forall \  i,j \in S, 
\end{equation}
In simple terms, one constructs the Boys-ADT matrix $U(R)$ by diagonalizing
the electronic position operator within $S$ at each nuclear position. 


Finally, note that before we can diagonalize Eq. \ref{eq:H_exact_diab},  we must enforce a phase-convention for the diabatic states without which the Hamiltonian is not well-defined.  To that end, below we enforce the parallel transport condition on electronic wavefunctions (which is always possible in one dimension). In other words, we match the phases of the diabatic wavefunction along $R$ so that $\left<\eta| d/dR|\eta\right> =0$. As a side note, for the case of adiabatic wavefunctions in Eq. \ref{eq:BH_vibronic}, we must also insist on the analogous smooth phase condition, $\left<\Phi|d/dR|\Phi\right> =0$.
Thereafter, the $K N_R \times K N_R$ diabatic Hamiltonian matrix can be diagonalized to obtain BO vibrational energies.

Now, formally, the eigenvalues of  Eq. \ref{eq:BH_vibronic} and Eq. \ref{eq:H_exact_diab} must be identical. That being said, as noted above, the intuition behind the development of a diabatic basis is to ignore any residual derivative coupling and diagonalize the much more inexpensive (but presumably less accurate) matrix
{\begin{equation}
    (\hat{H})_{ij} \approx  \left( \hat{\tilde{H}}\right)_{ij}  = \frac{\hat{P^2}}{2M}\delta_{ij} + \hat{\tilde{V}}_{ij}
\end{equation}}
while including as few electronic states as possible.  Very often this approach works, but the question of how many states are required for how much accuracy often complicates realistic {\em ab initio} calculations; formally, we must include all electronic states for the exact answer.
This concludes our brief review of how to extract vibronic energies within a BH framework.


\subsection{A Phase Space Diabatic Subspace Approach to Multireference Problems}
{The premise of the phase space electronic structure approach is that one can recover more accuracy than the BH framework by including nuclear-electronic correlation into adiabatic electronic structure calculations. Unlike Eq. \ref{eq:adiab} above, we parameterize electronic wavefunctions by both nuclear position $R$ and momentum $P$ via an additional electronic operator $\hat{\Gamma}$.} 
While many more details can be found in Refs. \citenum{bian_phase-space-way_2025,bian_PS-vibration_2025}, at the end of the day, for one nucleus in 1D, the PS Hamiltonian is of the form:
\begin{equation}\label{eq:PS_adiab}
    \hat H_{W}^{PS}(R,P) = \frac{(P - i\hbar\hat \Gamma)^2}{2M} + \hat H_{el}(R)
\end{equation}
Here, $P$ and $R$ are parameters in nuclear phase space and $\hat \Gamma$ is a nuclear position dependent electronic operator that aims to approximate the derivative coupling in Eq. \ref{eq:BH_vibronic} -- but which is a true one-electron operator rather than a $\partial/\partial R$ response matrix (see below). {Eq. \ref{eq:PS_adiab} can be written in a similar vein to Eq. \ref{eq:BH_vibronic_expanded} by expanding the square:}

{
\begin{equation}\label{eq:PS_adiab_expanded}
    \begin{aligned}
    \hat{H}^{PS}_{W}(R,P) &= \frac{P^2}{2M} +\left( \hat{H}_{el}(R) -i\hbar \frac{P\cdot \hat{\Gamma}}{M} - \hbar^2\frac{\hat{\Gamma}^2}{2M}\right)\\
         &= \frac{P^2}{2M} + \hat{H}_{W,el}^{PS}(R,P)
    \end{aligned}
\end{equation}}
{Although Eq. \ref{eq:PS_adiab_expanded} might appear to resemble Eq. \ref{eq:BH_vibronic_expanded}, 
there are two clear differences.
First, in Eq. \ref{eq:PS_adiab_expanded} we find 
the Weyl symbol $P^2/2M$ for the nuclear kinetic energy; this symbol is not an operator and does not operate on the electronic space as does $\hat {P}^2/2M$ in Eq. \ref{eq:BH_vibronic_expanded}. 
Second, 
while both 
$\hat{H}_{W,el}^{PS}$ and
$\hat{H}_{el}$
operate on the electronic space,
only
the former contains derivative coupling-like terms (and thus can capture more physics than the latter).
}

{With this background in mind, we postulate that, just as within BH theory, one can generate a strong electronic representation by diagonalizing
$\hat{H}_{W,el}^{PS}$ (instead of
$\hat{H}_{el}$). Let us define $\hat{U}_{W,el}(R,P)$ 
as the unitary transformation
that diagonalizes the PS electronic Hamiltonian Eq. \ref{eq:PS_adiab_expanded} at each nuclear phase space point $(R,P)$.}
\begin{equation}\label{eq:PS_diag}
    \hat{H}_{W,el}^{PS}(R,P) = \hat{U}_{W,el}(R,P)\hat{\Lambda}_{W}^{PS}(R,P)\hat{U}^{\dagger}_{W,el}(R,P)
\end{equation}
{Similar to the Born-Huang adiabatic approach, $\hat{\Lambda}^{PS}_W$ is a diagonal matrix whose elements indicate the electronic energy at each point in nuclear phase space $(R,P)$, while the eigenvectors (which are columns of $\hat{U}_{W,el}$) are denoted as PS adiabatic states.}

\subsubsection{Adiabatic (Single-State) Framework}

{When performing PS electronic structure calculations, one of the difficulties is that one generates electronic states for different nuclear $R$ and $P$ parameters--but in truth, the nuclei are quantum mechanical and $R$ and $P$  quantum observables that do not commute. Thus, if we wish to compute vibrational energies or other observables associated with nuclear degrees of freedom, one must ``requantize'' the nuclei after the diagonalization in Eq. \ref{eq:PS_diag}. How one performs such a requantization will necessarily depend on how many electronic  states one wishes to includes in the calculation.}

{
 Previous work in phase space electronic structure theory has so far focused on the most simple case, i.e. the case 
of a vibration that depend only a single electronic surface.
\cite{bian_PS-vibration_2025} 
For such a problem,
one can generate a vibrational Hamiltonian by interpreting $R$ and $P$ in Eq. \ref{eq:PS_adiab} as Weyl symbols in phase space and then invoking an inverse-Weyl transform on $\hat{\Lambda}^{PS}_{W}$. As a reminder, for any observable $\hat{O}$, the Wigner-Weyl invertible transformations are of the form:}

{\begin{equation}\label{eq:wigner_trans}
    \hat{O}_W(R,P) = \int dR' \bra{R+\frac{R'}{2}}\hat{O}\ket{R-\frac{R'}{2}}e^{-\frac{i}{\hbar}R'\cdot P}
\end{equation}}

\begin{equation}\label{eq:inverse_Weyl}
    \bra{R}\hat{{O}}\ket{R'} = \int \frac{dP}{2\pi \hbar} e^{(i/\hbar) P\cdot (R-R')}\hat{O}_W\left(\frac{R+R'}{2},P\right)
\end{equation}
{
Thus, in the context of the ground state phase space energy, $\hat \Lambda_W^{PS}$, one can generate the matrix elements of a vibrational (nuclear) Hamiltonian in position space as follows:
\begin{equation}\label{eq:ss_PS_inverse_Weyl}
    \bra{R,0}\hat{\Lambda}\ket{R',0} = \int \frac{dP}{2\pi \hbar} e^{(i/\hbar) P\cdot (R-R')}\left(\hat{\Lambda}^{PS}_W\right)_{00}\left(\frac{R+R'}{2},P\right)
\end{equation}
}
{
Ref. \citenum{bian_PS-vibration_2025} demonstrates that, 
if one diagonalizes the nuclear operator $\left(\hat{\Lambda}_{W}^{PS}\right)_{00}(R,R')$,
one can recover very accurate vibrational energies for a single surface well-separated from other surfaces}; in particular, if one considers artificial Hamiltonians when the electronic to nucleus mass ratio is reduced, such a PS approach can vastly outperform BO theory. Furthermore, according to Ref. \citenum{wu_exact-vibration_2025}, the approach above can be extrapolated to the exact limit following Littlejohn-Flynn\cite{littlejohn-flynn_perturbation-theory_1991} theory if one seeks to understand the approximation in terms of an infinite expansion. 

\subsubsection{Diabatic (Multi-State) Framework}

In what follows, our interest is to construct a multi-state PS Hamiltonian and extend the single surface result of Ref. \citenum{bian_PS-vibration_2025}. To that end, in analogy with the BH approach above, our approach will be to generate phase space adiabatic eigenstates $\left\{ \phi_k(r;R,P) \right\}$ and then
apply an ADT at each point in phase space to obtain diabatic phase space wavefunctions:
\begin{equation}
    \Xi_i(r;R,P)=\sum_k \phi_k(r;R,P)(U(R,P))_{ki}
\end{equation}

The corresponding diabatic Hamiltonian is similarly generated via the ADT to obtain PS diabatic energies and couplings in a phase space framework.

\begin{equation}
    \left(H_W^{PS}\right)_{ij}(R,P) = \sum_{k,l\in S} \big( \hat{U}_W(R,P)\big)_{ik}\big( \Lambda^{PS}_W\big)_{kl}\left(\hat{U}_W(R,P)\right)_{lj}
\end{equation}
As stated above, $\left(H_W^{PS}\right)$ is a $K N_R \times K N_R$ matrix. To requantize this matrix, we next perform the inverse-Weyl transform on each $N_R \times N_R$ Block of $H_W^{PS}$:
{
\begin{equation}
    \bra{R,i}\hat{H}^{PS}\ket{R',j} = \int \frac{dP}{2\pi \hbar} e^{(i/\hbar) P\cdot (R-R')}\left(\hat{H}^{PS}_W\right)_{ij}\left(\frac{R+R'}{2},P\right)
\end{equation}
The final vibrational energies are then determined by diagonalizing $\hat{H}$ in the $\ket{R,j}$ basis. For example, if we have $N_R$ grid points and $\hat{H}_{00}$  is an $N\times N$ operator, 
for a two state calculation, we will need to diagonalize a $2N \times 2N$ Hamiltonian of the form:
\begin{eqnarray}
    \hat H^{PS} = \begin{pmatrix}
        \hat{H}_{00} & \hat{H}_{01}\\
        \hat{H}_{10} & \hat{H}_{11}
    \end{pmatrix}
\end{eqnarray}
}

\section{Results}\label{Sec:III}
\subsection{Model}
We have run the calculations described above on a well known model for electron transfer and proton-coupled electron transfer developed by Shin and Metiu \cite{shin_nonadiabatic_1995,shin_multiple_1996}. For this 1D 1-electron 3-ion model, the edge ions are fixed in place, while the center ion is allowed to move. This scenario corresponds to the simplest possible Hamiltonian that describes adiabatic/nonadiabatic electron transfer while including a full description of both the electronic and nuclear degrees of freedom. For this reason, a host of theorists have studied the Shin-Metiu model, especially in the context of exact factorization\cite{neepa_XF_2019,agostini_XF-shin-metiu_2014}.

The Hamiltonian is parameterized as follows:
\begin{equation}\label{eq:H_exact}
    H = \frac{\hat{P}^2}{2M} + \frac{\hat{p}^2}{2m}+ V(R,r;L,R_f,C)
\end{equation}
Here, $L$ denotes the distance between edge ions, $R_f$ and $C$ are screening parameters for electron-ion interactions, $M$ is the mobile nuclear mass, and $m$ is the electron mass, which will be fixed to 1 atomic unit for all calculations. The nuclear-electronic potential has the following form:
\begin{equation}\label{eq:shin_metiu-potential}
    V(r,R) = \frac{1}{|\frac{L}{2}-R|} + \frac{1}{|-\frac{L}{2}-R|} - \frac{\mathrm{erf}\left(\frac{|R-r|}{R_f}\right)}{|R-r|}- \frac{\mathrm{erf}\left(\frac{\frac{L}{2}-r}{C}\right)}{|\frac{L}{2}-r|}- \frac{\mathrm{erf}\left(\frac{-\frac{L}{2}-r}{C}\right)}{|-\frac{L}{2}-r|}
\end{equation}
Changing the screening parameters produces a smooth transition between adiabatic and strongly nonadiabatic electronic surfaces. For this work, we have fixed the mobile ion screening parameter ($R_f$) to 5 {bohr}$^{-1}$ and vary the ion screening parameter ($C$) from 2 {bohr}$^{-1}$ to 10 {bohr}$^{-1}$. The fixed-ion distance was set to $L=20$ {bohr}. All vibrational energy calculations used a nuclear grid size of $N_R=151$ grid points along $R \in [-9,9]$ {bohr} and an electronic grid size of $N_r=151$ grid points along $r\in[-22,22]$ {bohr}. 

As far as the exact calculations are concerned, we diagonalized the $N_R N_r \times N_R N_r$ Hamiltonian matrix in Eq. \ref{eq:H_exact} in the position grid basis. Here, we applied a Davidson algorithm\cite{davidson_algorithm_1975} to converge the lowest two roots with a convergence threshold of  $10^{-12}$ Hartree for the energy deviation and $10^{-6}$ Hartree  for the residual.

Next, for the PS calculations, we use a momentum grid size of $N_P=151$ along $P\in [-\pi/\Delta R + \pi /(N_P \Delta R),\dots,\pi/\Delta R - \pi/(N_P \Delta R)]$ in atomic units based on the conjugate Fourier transform and  we choose 
\begin{eqnarray}\label{eq:gamma_simple}
   \hat \Gamma=\hat{p}/i\hbar. 
\end{eqnarray}
 We note that, for those familiar with our previous PS-electronic structure theory calculations\cite{bian_PS-vibration_2025}, this choice of $\hat \Gamma$ may appear unconventional insofar as $\hat{\Gamma}$ is usually distributed over all atomic sites \cite{tao_basis-free_nodate}. Indeed, the question of how to pick $\hat{\Gamma}$ when some sites are fixed and immobile is not obvious, and our choice of $\hat \Gamma$ and its deficiencies will be discussed below (Sec. \ref{Sec:IV}) where we provide data with alternative, more localized choices of $\hat \Gamma$.
Lastly, with PS calculations, the question of the phases of the electronic states is crucial -- much more so than for BO states  -- because the wavefunctions are complex. To that end, for our PS calculations, after diagonalizing $H^{PS}_{W,el}$, our first step is to diabatize at each point in phase space via a Boys/GMH diabatization\cite{cave_GMH_1996,cave_GMH_1997,boys1966quantum} (Eq. \ref{eq:boys-condition})  and then second we enforce a 1D parallel transport condition on the wavefunction phases. We enforce parallel transport in the following manner: i) We align the phases along the $P=0$ line extending forward (and backward) from $R=0$. ii) For each nuclear position $R$, we align the phases along the $P$ direction extending forward (and backward) from $P=0$. After these steps, we implement the inverse-Weyl transform in Eq. \ref{eq:inverse_Weyl} and diagonalize the $K N_R \times K N_R$ diabatic vibrational Hamiltonian.

Finally, for BH calculations, there are two options: (i) We can diagonalize $H_{BH}$ along one surface (the BO approximation) as shown in Eq. \ref{eq:BO_vibronic}; (ii) We can diagonalize the corresponding $\hat{H}_{BH}$ generated by Boys/GMH diabatization when working within the BH framework.

\subsection{Multistate Crossing in Shin-Metiu Surfaces}
\begin{figure}[ht]
    \centering
    \includegraphics[width=0.85\linewidth]{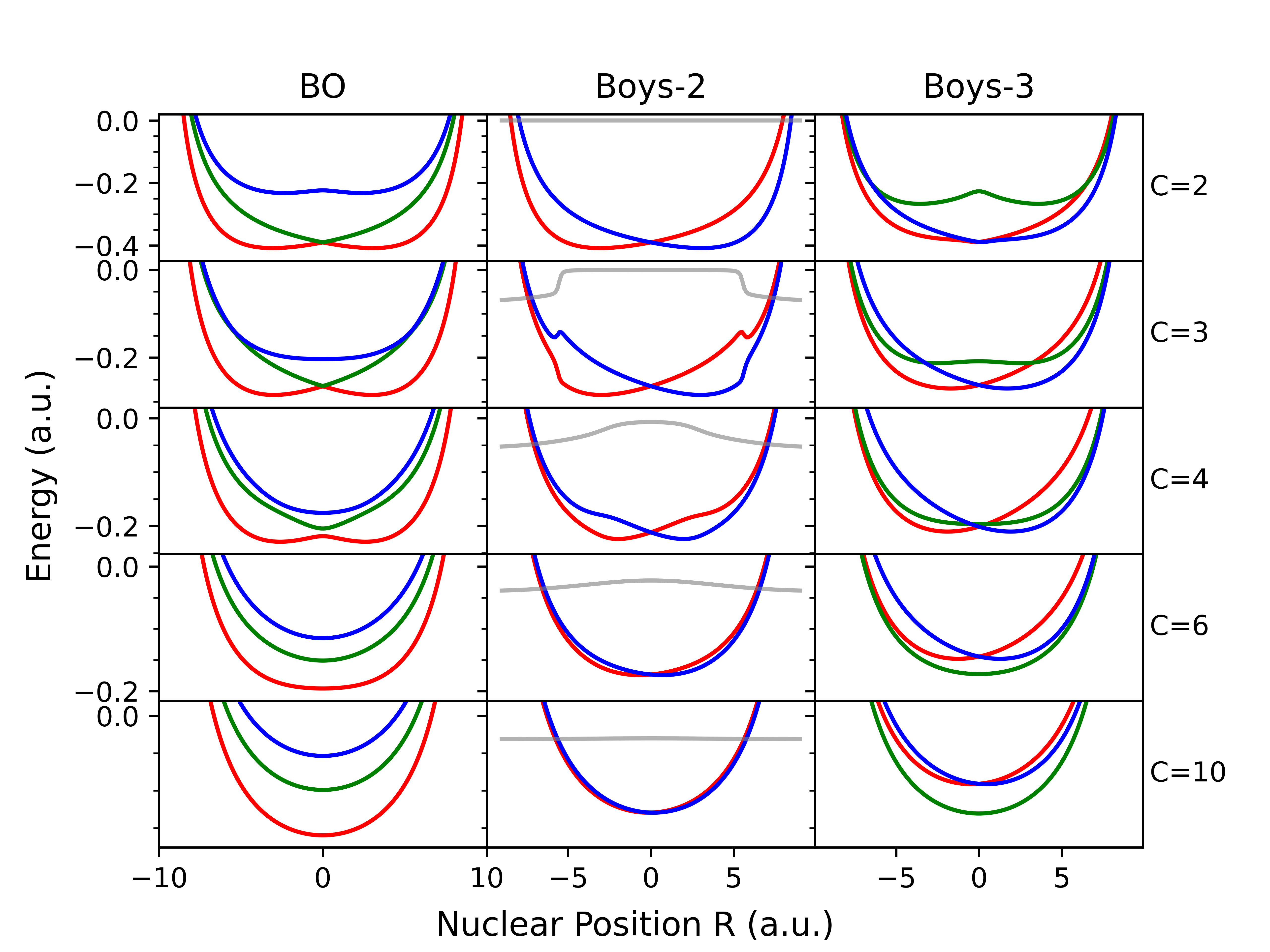}
    \caption{Born-Oppenheimer Adiabatic (Left), 2-State Boys (Center), and 3-State Boys (Right) surfaces for various external ion screening constants, denoted on the right of each row. Boys Surfaces are colored based on diabat (Left is Red, Blue is right, Green is Center). For 2-State Boys, diabatic coupling is shown in grey, and couplings are not shown for the 3-State Boys for clarity. Note that, due to the third state becoming an intruder at C=3, two-state Boys becomes unstable. Three state boys alleviates the instability at intermediate $C$ but becomes unstable when the third state becomes well separated as seen in the C=2 case.}
    \label{fig:bo_surfs}
\end{figure}
As noted by Shin and Metiu\cite{shin_nonadiabatic_1995}, several different regimes can emerge (depending on parameters) when using Eq. \ref{eq:shin_metiu-potential} to study nuclear-electronic correlation. Consider the adiabatic electronic surfaces at various ion screening constants $C$ in Figure \ref{fig:bo_surfs}. Depending on the choice of the ion screening constant $C$, two or three electronic surfaces become important for describing the low-energy physics of the problem, and the change from two to three encapsulates the transition between different forms of electron transfer processes. 
For large $C$ ($C>6$ a.u.), the adiabatic surfaces are well separated; the fixed ions do not significantly interact with the electron and the mobile-nucleus and electron behave more as a hydrogen atom trapped in a well. 
For intermediary $C$ ($6>C>5$ a.u.), the lower two surfaces start to approach each other in energy. The fixed ion-electron interaction creates a weak bond between the mobile nucleus and the nearest fixed ion; this regime corresponds to hydrogen atom transfer and proton coupled electron transfer (HAT/PCET). 
For small $C$ ($5>C>3$ a.u.), a third adiabatic electronic surface crosses with the second adiabatic surface (acting as an intruder state), and as a consequence, 2-state diabatization methods become unstable.
Finally, for very small $C$ ($3>C>2$ a.u.), the fixed ion wells become so deep that the system is better described as two separate deep atomic wells; the mobile nucleus breaks the symmetry but does not attract any electronic density.  Because the nature of the ground state is incredibly sensitive to the position of the mobile nucleus, the
electron transfer is entirely nonlocal.
This limit can be best described as extremely electronically nonadiabatic PCET\cite{migliore-beratan_pcet-marcus_2014,sharon_nonadiabatic-pcet_2005}. Note that, in this regime, if one considers the third electronic surface, one finds that the latter becomes well separated from the lower two surfaces but exhibits its own avoided crossing with the fourth electronic surface. 

As a practical matter, one often discusses the parameter regimes above in terms of the degree of nonadiabaticity of the problem at hand. To estimate such a degree, we will choose to use the ratio of Marcus parameters as the metric for electronic nonadiabaticity. Specifically, given two diabats, the ratio of the reorganization energy $E_R$ and diabatic coupling $V_{DA}$ at the crossing point is a standard measure of the degree at which the upper adiabatic surface is relevant to the electron transfer process. If $E_R \ll V_{DA}$ ($E_R/V_{DA} \ll 1$), diabatic coupling strongly mixes the two diabats, leading to well separated adiabatic surfaces. Alternatively, if $V_{DA} \ll E_R$ ($E_R/V_{DA} \gg 1$), the two diabats are very weakly coupled,  mixing only at the crossing point and the associated adiabats approach a  nearly trivial crossing. 

In summary, by tuning the model parameters in the Shin-Metiu paper, one can study adiabatic vibrations, hydrogen atom transfer, proton coupled electron transfer, and extremely nonadiabatic electron transfer, while smoothly transitioning between these phenomena. However, depending on the screening constant, one would prefer diabatization with either two or three states; there is no obvious diabatization scheme that works for all parameter regimes. To that end, below we will compute vibronic energies using both 2-state and 3-state Boys-diabats.

\subsection{Vibrational Energy Gap in Shin-Metiu Systems}
\begin{figure}[t]
    \centering
    \includegraphics[width=\linewidth]{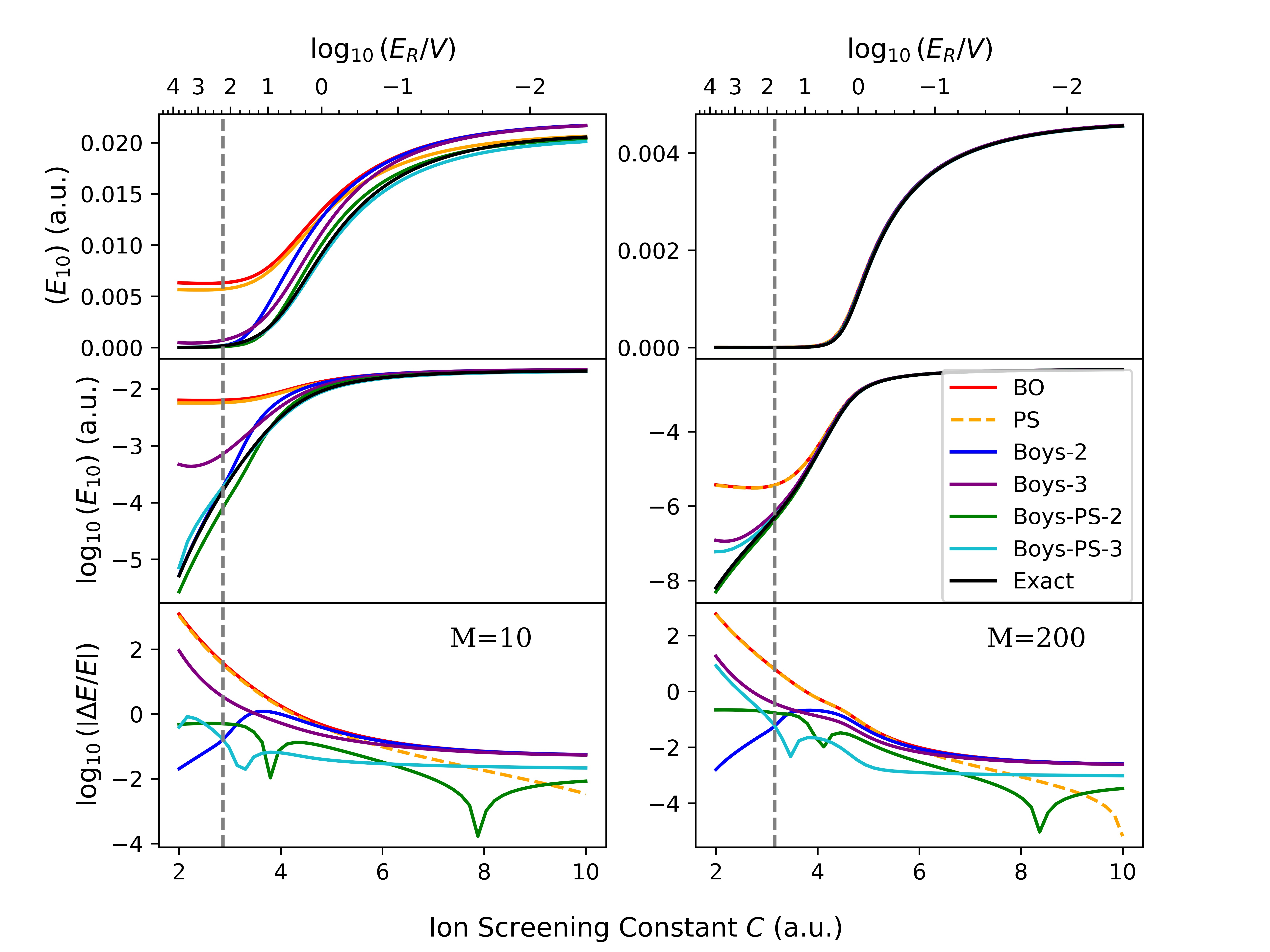}
    \caption{Lowest vibrational energy gap  on a linear scale (top), on a log scale (middle), and the relative absolute error of the vibrational energy gap on a log scale (bottom) as a function of ion screening constant $C$ (bottom axis) or log of the nonadiabaticity parameter (top axes). We plot results for electron-mobile ion mass ratios of 10 (left) and 200 (right). Note that the dips present in PS results indicate when the calculated-energy crosses the exact energy. The point at which PS outperforms BH is indicated by the gray vertical dashed line. All PS calculations in these plots use $\hat \Gamma=\hat p/i\hbar$. In general, for $C > 3.5$ a.u., one finds that a Boys-PS strongly outperforms all other results. }
    \label{fig:main_results}
\end{figure}

To assess the methods described above, in Figure \ref{fig:main_results}, we have computed the vibronic energy gap between the lowest two vibrational states of the Shin-Metiu model at different values of $C$ reported on the lower x-axis.  On the upper x-axis, we list the value of the nonadiabatic parameter $E_R/V$ described in Sec. \ref{Sec:III}. 
Note that, as the ion screening constant decreases and  the degree of nonadiabaticity increases, 
the nonadiabaticity parameter rapidly increases can reach $10^3-10^4$ as we approach the regime of a trivial crossing.
To probe sensitivity to nonadiabatic effects, we perform several different calculations with a variable ratio of the nuclear to electronic mass ($M/m=10$ and $M/m=200$). {It is important to note that, for real proton coupled electron transfer, the mass ratio is closer to $M/m=2000$. However, the framework for electron transfer 
is the same as the frame work for proton transfer\cite{peters-proton_tranfsfer-2009,borgis-curveCrossing_proton-1996,kiefer-nonadiabaticPT-2004,sharon-PCET_rev-2010}. Thus, one would like to find a single method that can treat electron or proton transfer, in the adiabatic or nonadiabatic  limits, with variable mass ratios. Note that a smaller mass mass ratio also reduces the electronic gap, which then allows us to explore new physics which should be important for quantifying spin-dependent electron transfer.}

Let us  consider first the BH results. 
According to Figure \ref{fig:main_results}, not surprisingly,  single-state BO gives satisfactory results only for $C>6$ a.u; very poor results arise at  intermediate or otherwise {smaller} screening constants. 
Interestingly, if  we invoke diabatization, we find that 2-state Boys is  fairly accurate (both in the large $C$ and especially in the small $C$ limit). That being said, the method is unstable and gives markedly worse results in the region of $3.5 < C < 5$ a.u. because of the three state intruder problem. By contrast, the 3-state Boys maintains smooth surfaces and couplings everywhere, but loses a great deal of accuracy in the nonadiabatic, small $C$ limit.

Second, consider the PS results. 
In the range, $3.25 < C < 10$ a.u., we observe that both the two and three-state PS vibronic energy gaps consistently outperform all possible BH vibronic energy gaps, gaining up to two orders of magnitude of accuracy over BH. 
A three state diabatization appears smoother than a two state diabatization. 
That being said, in the range $2<C<3.25$ a.u. when the screening constant becomes sufficiently small, both  PS methods break down, and the 3-state model performs significantly worse than the 2-state model. Nevertheless, note this breakdown occurs only for  $E_R/V$ on the order of $10^2$; in short,  
Boys-PS is stable within the range of adiabatic to partially nonadiabatic surfaces, and breaks down only in the strong nonadiabatic limit.

\subsection{Position and Momentum Transition Moments in Shin-Metiu Systems}

{To complement the results above, we have also computed various other  non-energy matrix elements between the lowest two vibronic states. Nuclear operator matrix elements are obtained by evaluating $\langle \hat{O}\rangle_{mn} = \sum_{ij} \langle\chi_i^m| \hat{O} |\chi_j^n\rangle$ where $\chi_i^m(R)$ is the portion of the $m$th nuclear (vibrational) wavefunction associated with electronic state $i$. For electronic observables, the matrix elements between phase space electronic states $\ket{\phi_i(R,P)}$ are initially evaluated at each point in phase space.
\begin{equation}
    \left(o_W\right)_{ij}(R,P) = \bra{\phi_i(R,P)}\hat{o}\ket{\phi_j(R,P)}
\end{equation}
Thereafter, the resulting matrix elements $\left(o_W\right)_{ij}(R,P)$ are inverse-Weyl transformed into position space $(o)_{ij}(R,R')$, before averaging with the relevant nuclear wavefunctions:
\begin{equation}
    \langle\hat{o}\rangle_{mn} = \sum_{ij} \int dR \left(\chi_i^m(R)\right)^\dagger \hat{o}_{ij}\chi_j^n(R)
\end{equation}}

\begin{figure}
    \centering
    \includegraphics[width=\linewidth]{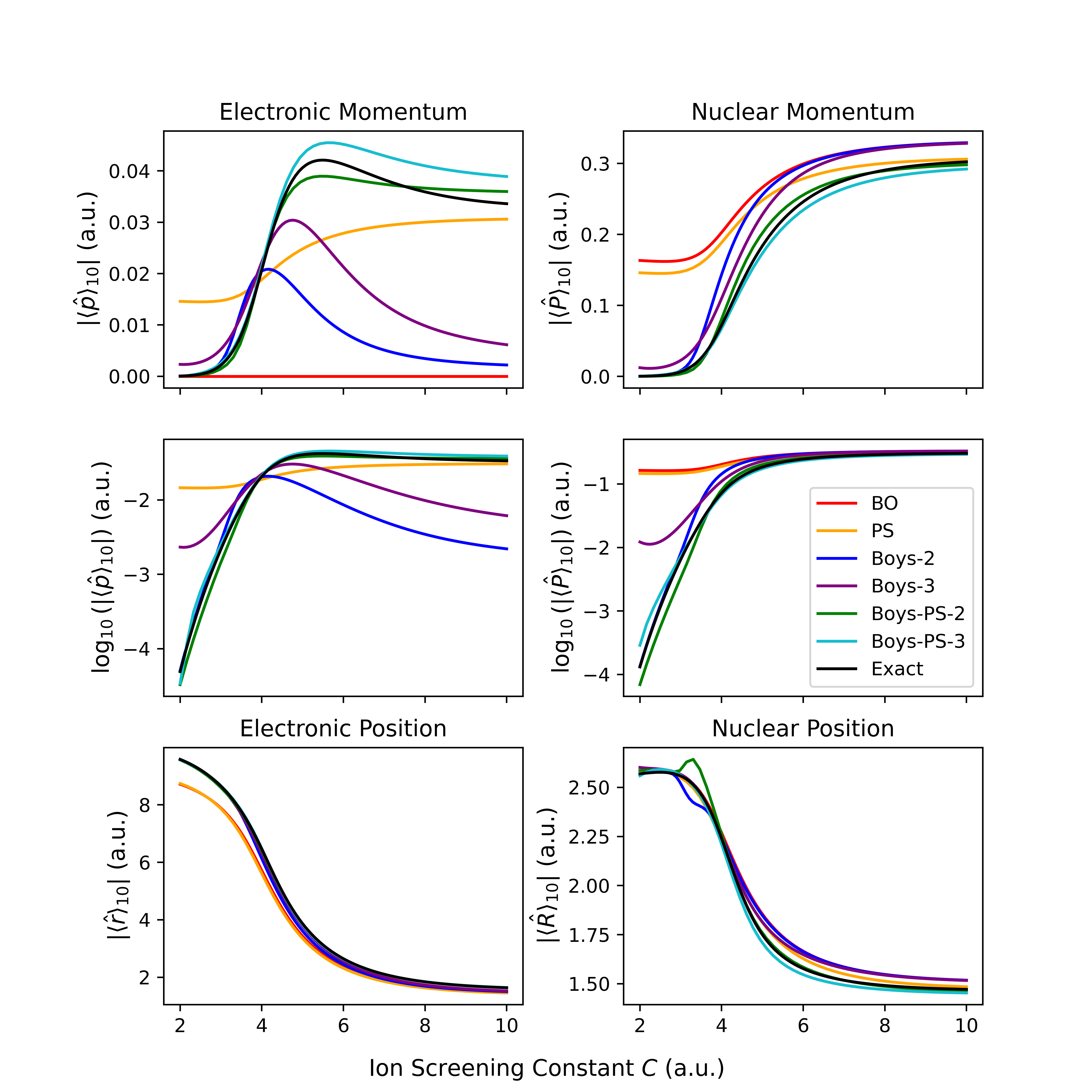}
    \caption{{Absolute magnitudes of various observables as calculated from a BH, phase space, or numerically exact approach as a function of ion screening constant $C$. All calculations were performed for a mass ratio of $M/m=10$. We plot the nuclear and electronic momentum on an absolute scale (top row) as well as a logarithm scale (middle row). Note that the large deviance observed in nuclear position around $C=3$ a.u. arises from the instability of a 2-state diabatization model. Similar plots for a mass ratio of $M/m=200$ can be found in the Supplemental Information. The intruder state problem appears in nuclear position $\langle R\rangle _{01}$ at $C=3$. Otherwise, for all observables, note that a multi-state phase space electronic structure approach almost always outperforms a BH approach. }}
    \label{fig:M10-MatElem}
\end{figure}

\begin{figure}
    \centering
    \includegraphics[width=\linewidth]{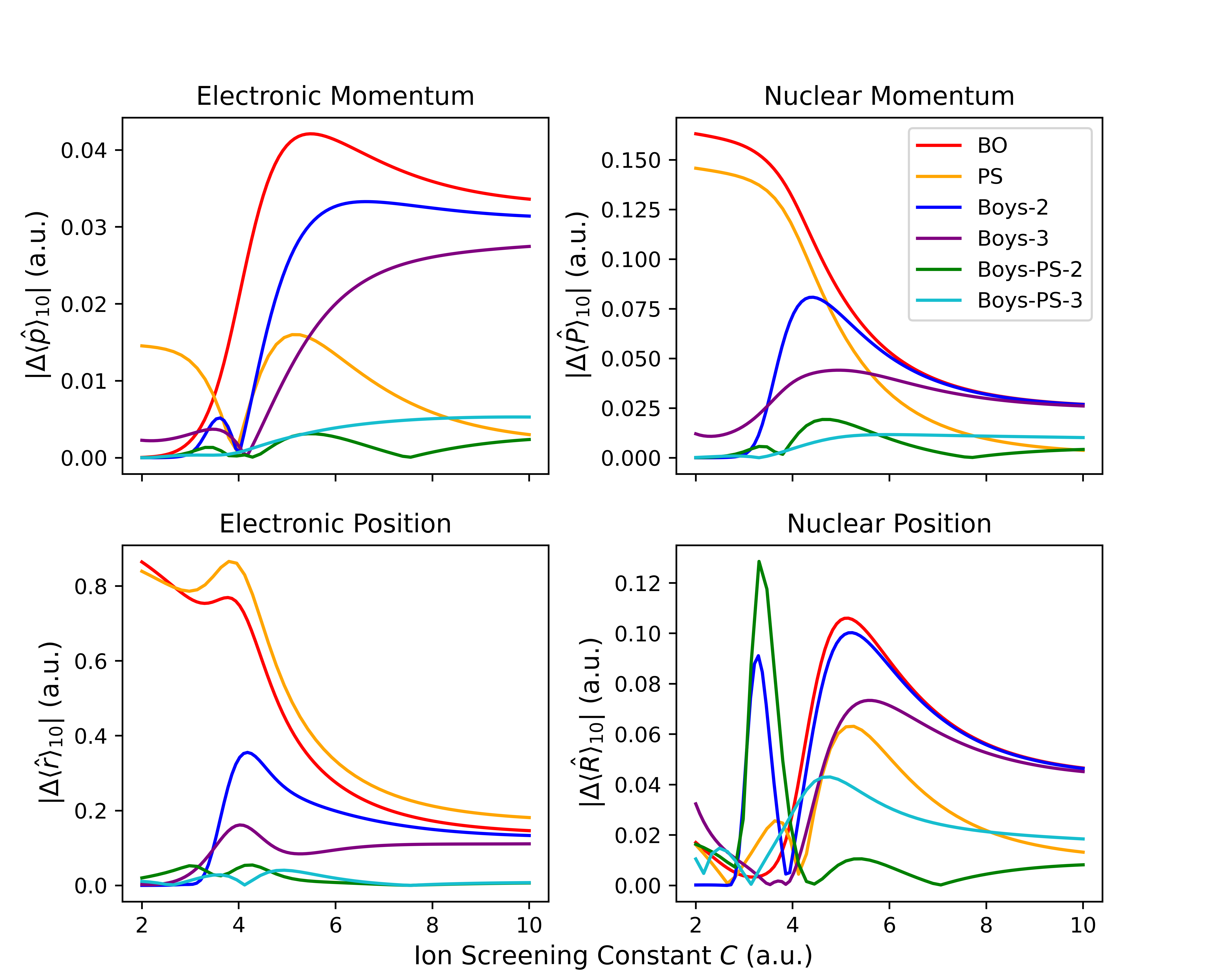}
    \caption{{Absolute errors ($\Delta$) of various observables as calculated from a BH and phase space approach compared to exact calculations as a function of ion screening constant $C$. All calculations were performed for a mass ratio of $M/m=10$. The spike in error observed in nuclear position around $C=3$ for the 2-state approaches is clearly visible. Similar plots for a mass ratio of $M/m=200$ can be found in the Supplemental Information.}}
    \label{fig:M10-absErrMatElem}
\end{figure}

\begin{figure}
    \centering
    \includegraphics[width=\linewidth]{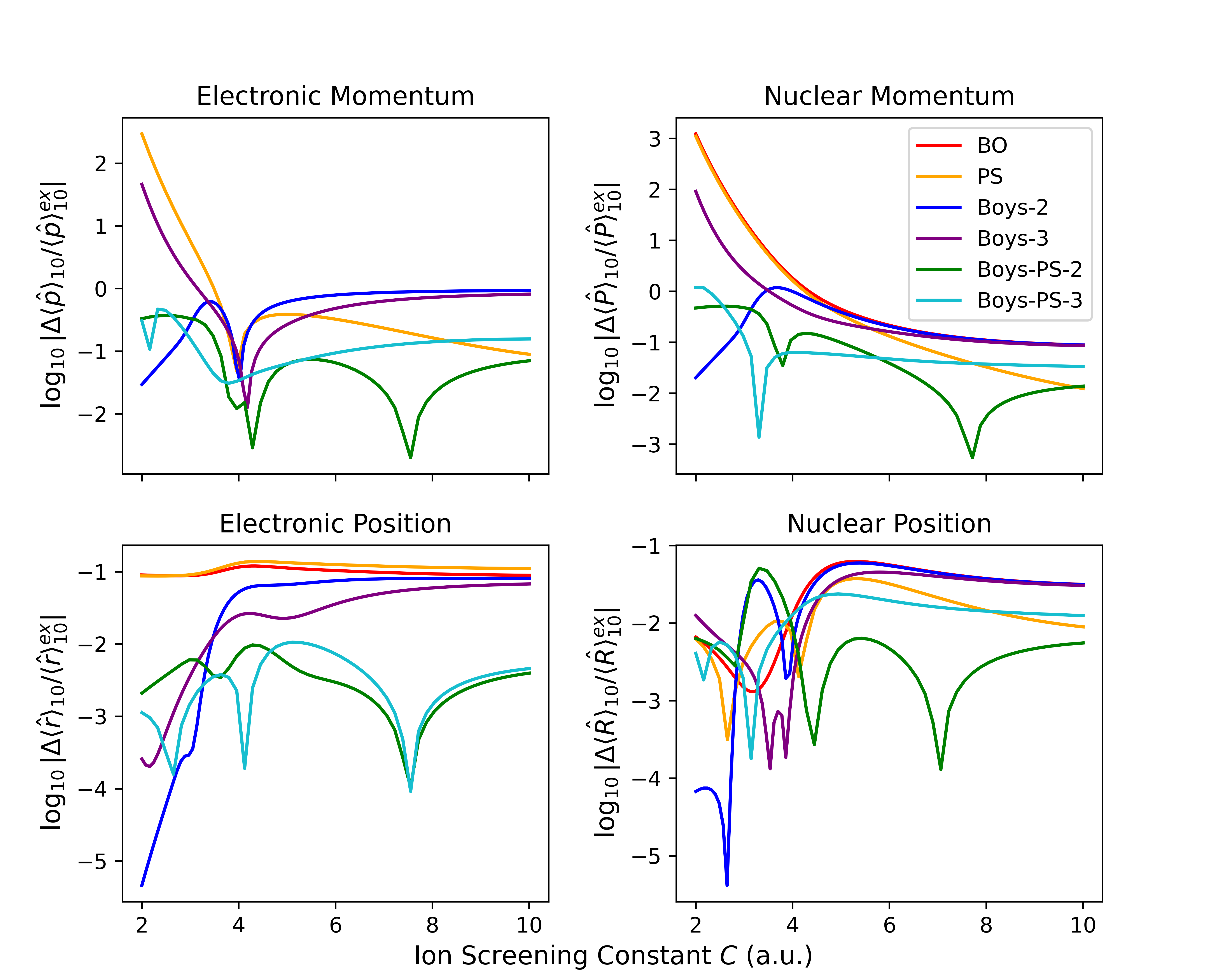}
    \caption{{Relative errors of various observables on a log scale as calculated from a BH and phase space approach compared to exact calculations as a function of ion screening constant $C$. All calculations were performed for a mass ratio of $M/m=10$. Similar plots for a mass ratio of $M/m=200$ can be found in the Supplemental Information. Note the very strong performance of the Boys-PS methods.}}
    \label{fig:M10-relErrMatElem}
\end{figure}

{As described in Ref. \citenum{bian_PS-vibration_2025}, the nuclear momentum operator $\hat{P}$ is replaced with $\hat{P} - i\hbar\hat{\Gamma}$ in phase space calculations, which represents the canonical-kinetic momentum correspondence. The results of these calculations are seen in Figures \ref{fig:M10-MatElem} - \ref{fig:M10-relErrMatElem}. Overall, three conclusions stand out. First, whereas BH theory ignores electronic momentum on a given electronic surface, so that $\langle \hat{p}\rangle_{01}=0$, no such assumption is made by PS theory and in fact, Boys-PS theory is  able to offer a very accurate estimate of this very important observable over all regimes (adiabatic and nonadiabatic).  Second, the other observable with sometimes large errors is the nuclear momentum, where we find that, just as for energy, Boys-PS theory is able to offer very strong results --- and provided $C>3$, these results are in fact optimal and strongly outperform BH theory. 
Third, the data for electronic and nuclear position roughly follow the same trends in accuracy with the only exception being nuclear position $\langle \hat{R}\rangle_{01}$ clearly shows the influence of the intruder state around $C=3$. 
Overall, the quantities calculated in Fig. \ref{fig:M10-relErrMatElem} again point to a PS advantage over BH theory.}



\section{Discussion}\label{Sec:IV}

The results above are highly encouraging, but necessarily raise several questions.

\subsection{Does PS really offer a more accurate electronic subspace than BH for low energy excitations?}

\begin{figure}[H]
    \centering
    \includegraphics[width=\linewidth]{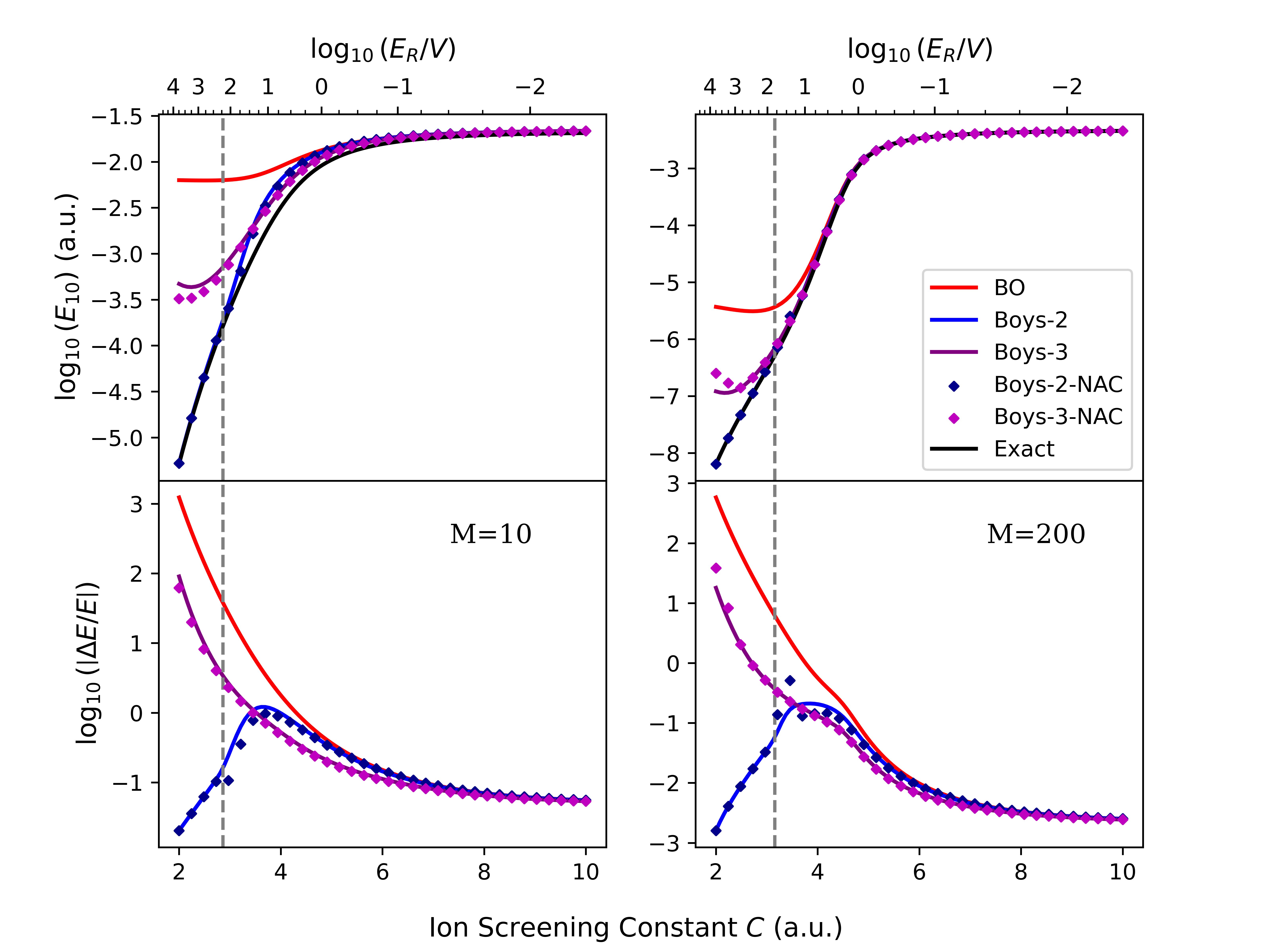}
    \caption{(Top) The vibrational energy gap with or without including nonadiabatic couplings in the diabatized subspace.  (Bottom) The relative absolute error in the vibrational energy gap for the same methods as above.  Note the minor instability in Boys-2-NAC calculations (blue, bottom right) near the vertical grey dashed line; this instability occurs due to the numerical instability of a 2-state model in this region caused by the presence of an intruder state.  Overall, from this data, one can infer that a Boys-PS approach performs well by including the effects of higher lying states--rather than by addressing the minimal residual derivative couplings between the two or three diabatic states included.}
    \label{fig:4}
\end{figure}

Above, we have argued that, in the limit of medium sized mass ratios, the PS approach for a subspace of strongly coupled states strongly outperforms the BH subspace. Now, a seasoned reader can also question whether such a PS advantage arises because the PS framework chooses a better set of electronic states than the BH framework, or because we have not optimally calculated vibronic energies within a BH framework (using Boys/GMH diabatization\cite{boys1966quantum,edminston-ruedenberg_localized_1963,cave_GMH_1996,cave_GMH_1997, subotnik_boys_2008}). For the strongest apples-to-apples comparison, let us now evaluate the vibronic energy gap using Eq. \ref{eq:H_exact_diab} and including the diabatic derivative couplings in the chosen subspace. For these calculations that are exact within a given BH subspace, note that our simulations are necessarily invariant between adiabatic and diabatic bases; there is no dependence at all on the use of Boys diabatization, etc.

Our results are shown in Fig. \ref{fig:4}. According to the data, including the derivative coupling
leads to a small improvement in the vibrational energy gap, but PS-Boys calculations with the same diabatic subspace still consistently outperform Boys-NAC for $C>3.25$ a.u.. 
{The same conclusion follows if we compare the results for transition moments ($\langle \hat{r}\rangle_{01}$ and $\langle \hat{p}\rangle_{01}$ ) relative to exact calculations, as in Fig. \ref{fig:4}.}
We may then conclude that the PS approach provides a much more accurate subspace of states for electron transfer than does BH. 

While this statement might at first appear surprising, note that  in the limit that all BH states are included, total momentum conservation is enforced through the derivative couplings;  truncation to a subset of Born-Huang surfaces leads to a violation of conservation of momentum (since derivative couplings outside of the subspace are neglected). By contrast, every single phase space surface automatically conserves the total linear and angular momentum, so one should not be surprised that the resulting subspace fundamentally captures more of the low-energy physics of the electron transfer problem. 

\begin{figure}[H]
    \centering
    \includegraphics[width=\linewidth]{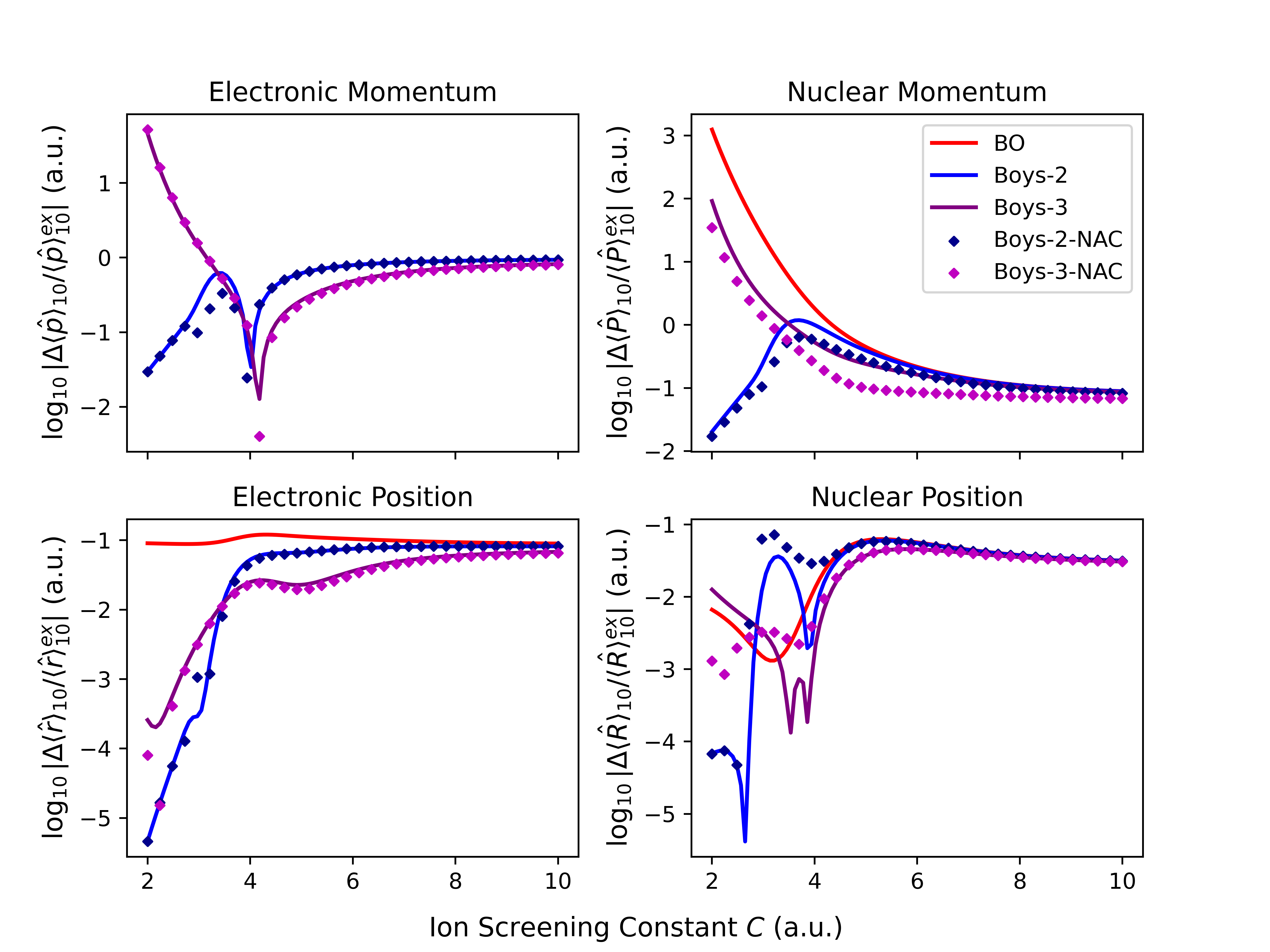}
    \caption{{Relative errors of various physical observables on a log scale using BH approaches with or without including nonadiabatic couplings as a function of ion screening constant $C$. All calculations were performed for a mass ratio of $M/m = 10$. Similar plots for a mass ratio of $M/m = 200$ can be found in the Supplemental Information.  As in Fig. \ref{fig:4}, note that including the nonadiabatic couplings within the two or three state subspace of interest does not much improve the results;  a PS approach almost always performs better. }  }
    \label{fig:m10-operators-nac}
\end{figure}

\subsection{Why does PS fail in the strongly nonadiabatic regime?}
\begin{figure}
    \centering
    \includegraphics[width=0.85\linewidth]{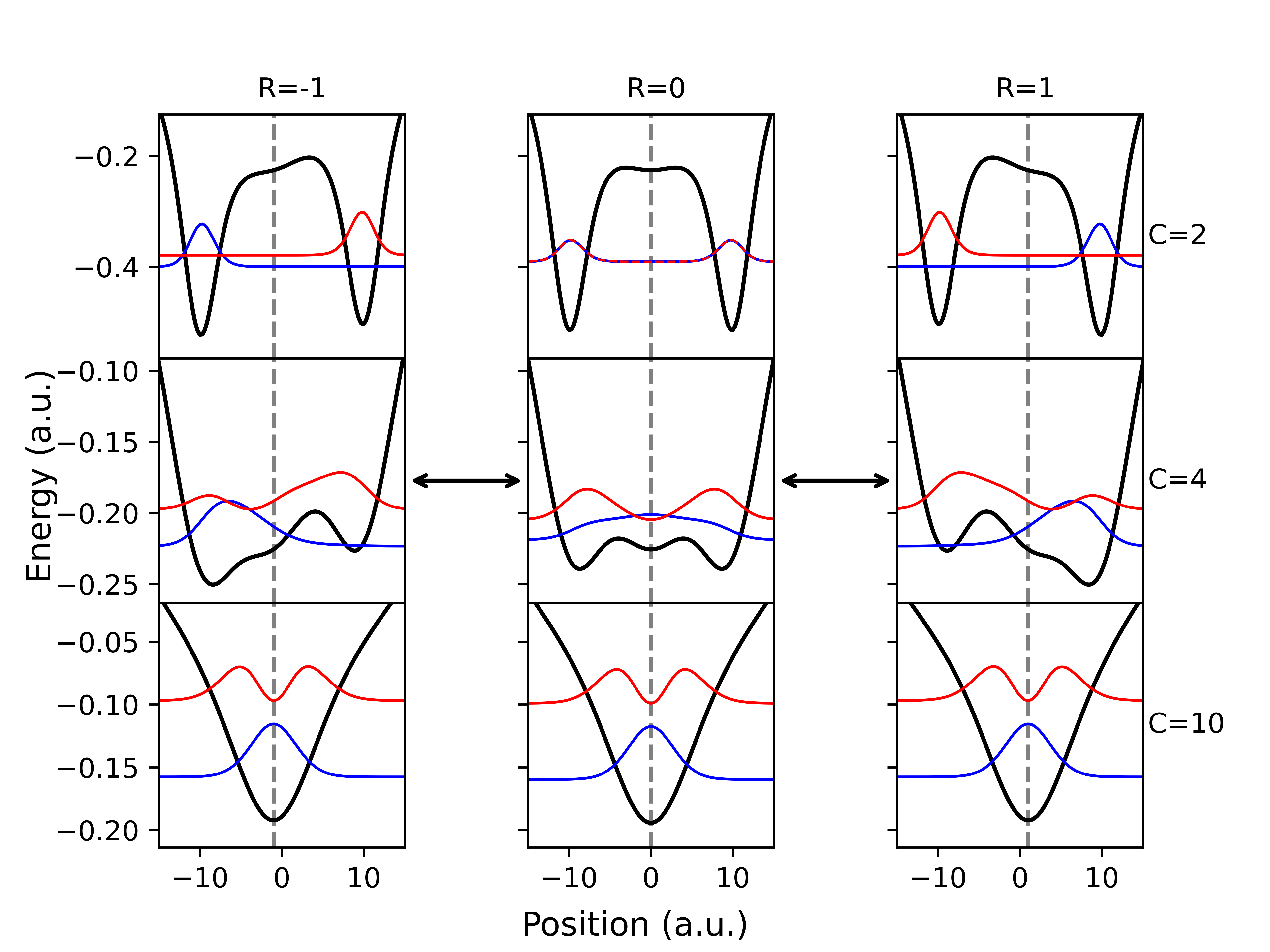}
    \caption{Ground (blue) and 1st excited (red) adiabatic electronic wavefunction probability distributions a function of electronic position $r$ for  various nuclear positions $R$ and screening constants $C$. The adiabatic potential felt by the electron is shown in black and the position of the mobile nucleus is shown with the gray vertical dashed line. Here, we plot data for $C=2$ a.u. (top), $C = 4$ a.u. (middle) and $C = 10$ a.u. (bottom) in order to model the transition from nonadiabatic to adiabatic ET.  Note that, for the middle and lower panels (where PS performs quite well), there is some electronic density at $r=0$ for $R=0$.  However, no such density is present in the nonadiabatic limit, $C= 2$ a.u. Note that, in this same limit, for $R=0$, the ground and excited adiabatic surfaces are near-identical, with density split across both fixed ions.}
    \label{fig:electron-dens}
\end{figure}

To answer this question, it is imperative to recall our choice of $\hat \Gamma$ in Eq. \ref{eq:gamma_simple}.
Consider first the adiabatic regime.
It is well known that the derivative coupling of a hydrogen atom (1 electron and one nucleus) is exactly $\hat d = \hat p/i\hbar$\cite{bian_PS-vibration_2025}; in other words, Eq. \ref{eq:gamma_simple} would be exact if we were to ignore the fixed ions.
Thus, physically, including such a $\hat \Gamma$ operator 
within a phase-space electronic structure picture corresponds to boosting the electron into the reference frame of the nucleus; at nonzero $P$, the electronic PS wavefunctions will also have a nonzero $\langle\hat p\rangle$. Now, in the adiabatic limit of the Shin-Metiu model, the electronic wavefunction is fully localized around the mobile nucleus; see Row 1 in Fig. \ref{fig:electron-dens}.  When the mobile nucleus moves, the electron responds strongly and stays with the nucleus.  Thus, the crude choice of $\hat \Gamma $ in Eq. \ref{eq:gamma_simple} is indeed quite appropriate.

Second, consider the strongly nonadiabatic limit where the situation is different.  See Row 3 in Fig. \ref{fig:electron-dens}. Here, the momentum of the mobile nucleus does not "drag" the electrons along with it.  Rather, the electronic rearrangement is entirely nonlocal and the character of the ground state is dictated merely by the small electric field produced the mobile nucleus. Unfortunately, the choice of $\hat \Gamma$  in Eq. \ref{eq:gamma_simple} erroneously applies a nonzero momentum globally, regardless of the presence of electronic density on the local atom, and thus  a deficiency in our definition of $\hat \Gamma$ leads to the breakdown of Boys-PS. 

\subsection{Can we choose an even more accurate $\hat{\Gamma}$?}

We have seen that our PS approach above fails in the nonadiabatic limit and we have argued that the root cause is the form of $\hat \Gamma$ in Eq. \ref{eq:gamma_simple}. To that end, one can certainly seek a better form of $\hat{\Gamma}$ and, following Refs. \cite{bian_PS-vibration_2025,tao_basis-free_nodate}, 
one simple  multi-nuclear form is \cite{tao_basis-free_nodate}:
\begin{equation}\label{eq:gamma-local}
    \hat \Gamma_A = \frac{1}{2i\hbar} (\hat{\theta}_A\hat{p} + \hat{p}\hat{\theta}_A)
\end{equation}
Here, $\hat \theta_A$ is a partition of unity (that satisfies $\sum_A \hat \theta_A = 1$). For example, one possible partition of unity is of the following form:
\begin{equation}\label{theta-func}
    \hat \theta_A = \theta_A(\hat{r}) = \frac{e^{-|\hat r - R_A|^2/\sigma^2}}{\sum_B e^{-|\hat r - R_B|^2/\sigma^2}}
\end{equation}
Here, $R_A$ is the location of nucleus $A$ and $\sigma$ controls the spread of the gaussian. 
The effect of including $\hat \theta_A$ in Eq. \ref{eq:gamma-local} is that the nuclear momentum is localized in a region near atom $A$ for all atoms $A$.

In principle, one can imagine two different limits. First, one can take the limit of small $\sigma$ (below we choose $\sigma = 4$ {bohr}$^{-1}$):
\begin{equation}\label{eq:gamma_small}
\hat{\Gamma} = \frac{1}{2i\hbar}\left( \frac{e^{-|\hat{r}-R|^2/\sigma^2}}{e^{-|\hat{r}-R|^2/\sigma^2} + e^{-|\hat{r}-\frac{L}{2}|^2/\sigma^2} + e^{-|\hat{r}+\frac{L}{2}|^2/\sigma^2}}\hat{p} + \hat{p}\frac{e^{-|\hat{r}-R|^2/\sigma^2}}{e^{-|\hat{r}-R|^2/\sigma^2} + e^{-|\hat{r}-\frac{L}{2}|^2/\sigma^2} + e^{-|\hat{r}+\frac{L}{2}|^2/\sigma^2}}  \right)
\end{equation}
This regime strongly localizes the electronic density to a nearby nucleus because $\theta_A$ looks like a heaviside characteristic function that is unity inside the domain of $A$ and zero outside. In the second limit, one takes $\sigma \rightarrow \infty$, which leads to 
\begin{eqnarray} 
\label{eq:gamma_one_third}
\hat{\Gamma} = \frac{1}{3}\frac{\hat{p}}{i\hbar}.
\end{eqnarray}
In this limit, the electron is split between all three nuclei (the one mobile and two fixed ions) for all geometries.  

\begin{figure}
    \centering
    \includegraphics[width=1.00\linewidth]{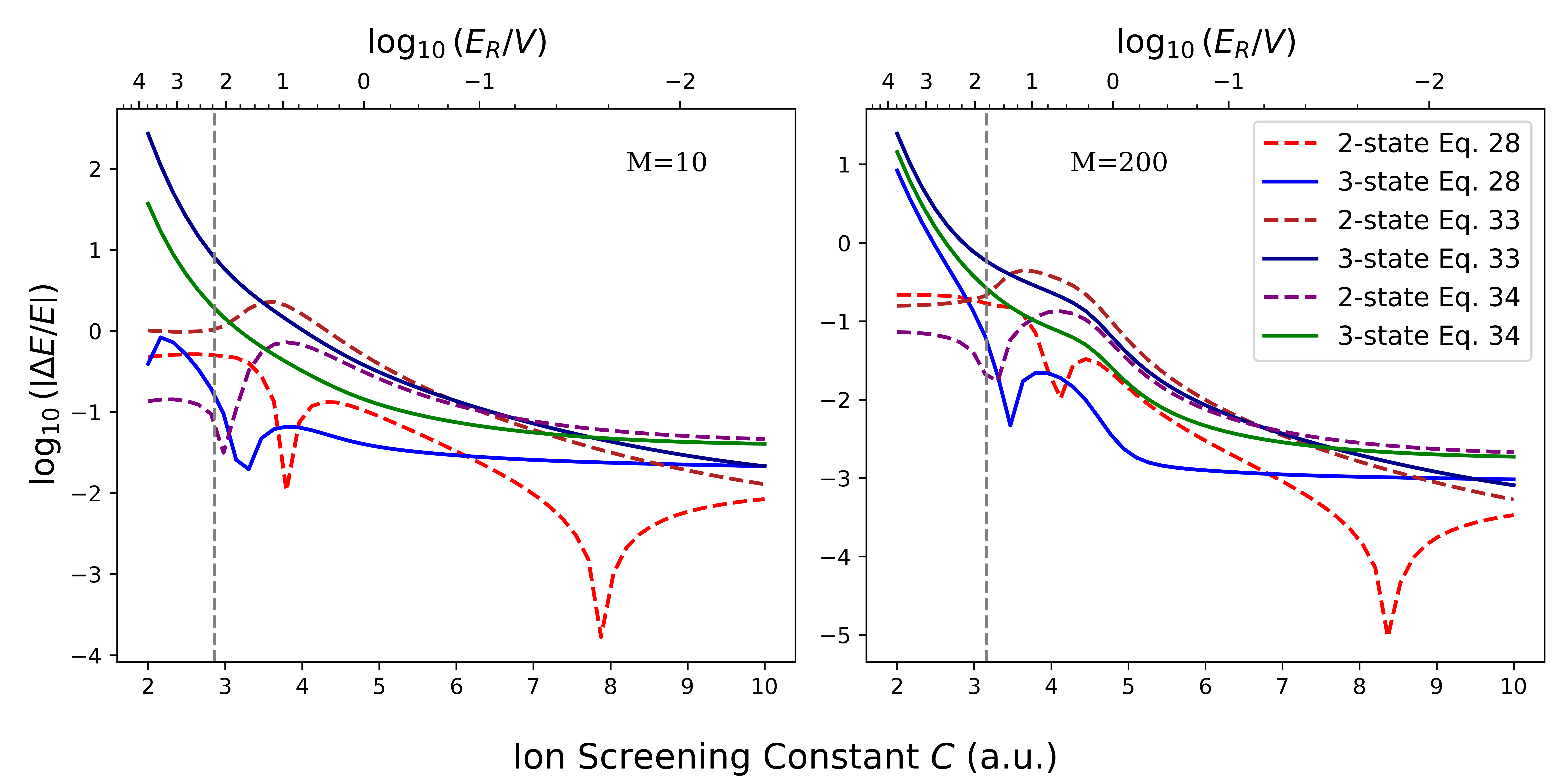}
    \caption{Relative absolute error of the vibrational energy gap on a log scale as a function of the ion screening constant $C$(bottom axis) or log ofthe nonadiabaticity parameter (top axes) with varying definitions of $\hat \Gamma$.  In particular, we implement  $\hat \Gamma$ according to Eqs. \ref{eq:gamma_small} and \ref{eq:gamma_one_third}. The vertical dashed line is the same as in Fig. \ref{fig:results-gammas}.  Overall, notice that the choice of Gamma in Eq. \ref{eq:gamma_simple} outperforms all other choices for reasons described in the text.  }
    \label{fig:results-gammas}
\end{figure}

In Figure \ref{fig:results-gammas}, we show results for both cases above (Eq. \ref{eq:gamma_small} and Eq. \ref{eq:gamma_one_third}).  Within the partially nonadiabatic regime (C= 4-8 a.u.), perhaps not surprisingly, 
both methods above perform worse than Eq. \ref{eq:gamma_simple} above.  After all,  in the partially nonadiabatic regime, there is weak to strong bond formation and nuclear motion drives a large amount of electronic rearrangement in regions between the fixed ion and mobile ion.  However, both choices of $\hat \Gamma$ lead to only a small portion of the electronic wavefunctions having a non-zero electronic momentum. 
Vice versa, in the strongly nonadiabatic regime (where we should find no density on the mobile ion), we do find  a modest
improvement to the data using Eq. \ref{eq:gamma_one_third}, which eliminates 2/3 of the spurious electronic momentum ascribed to the mobile ion by Eq. \ref{eq:gamma_simple}. 
Ironically, this state of affairs is not improved by Eq. \ref{eq:gamma_small} (in the local limit with sigma = 4) because, given the fast decay of exponentials, one will inevitably find that, no matter the value of $\sigma$, if A is the mobile ion, $\theta_A (R_A)$ is always very large (which again results in a spurious electronic momentum for the electronic density around the mobile nucleus).

The analysis above makes clear that, for a realistic Hamiltonian, one must necessarily weight the $\theta_A$ functions in Eq. \ref{theta-func} for optimal accuracy. For this reason, previous papers have used a function of the form 
\begin{eqnarray}
    \hat \theta_A = \frac{Q_Ae^{-|\hat r - R_A|^2/\sigma^2}}{\sum_B Q_Be^{-|\hat r - R_B|^2/\sigma^2}}
\end{eqnarray}
or 
\begin{eqnarray}
    \hat \theta_A = \frac{M_A e^{-|\hat r - R_A|^2/\sigma^2}}{\sum_B M_B e^{-|\hat r - R_B|^2/\sigma^2}}
\end{eqnarray}
For most chemical problems, these definitions are quite similar insofar as atomic mass $(M_A)$ and atomic charge $(Q_A)$ are roughly proportional. That being said, when atoms are fixed, or when one considers non-physical Hamiltonians, one can anticipate the need for a more accurate approach to $\hat{\Gamma}$.  Hopefully, for molecular or material problems, where the atomic densities can usually be well described without calculating Born effective charges, our hope is that a simple model of $\hat{\Gamma}$ can be effective. Further research in this area will be needed in the future. 


\section{Conclusions and Future Directions: Degenerate States and Spin}\label{Sec:V}

We have demonstrated that employing a phase space electronic structure framework can provides a subspace of diabatic states that describe electron transfer processes with consistently more accurate results than traditional diabatization in a Born-Huang picture, so long as we are not in the strongly nonadiabatic regime. While our results are based on the Shin-Metiu model\cite{shin_nonadiabatic_1995}, we believe the analysis above is likely robust insofar as the interpretation is fairly straightforward.
Looking forward, there is clearly still some work to be done as far as smoothly extrapolating our results into the strongly nonadiabatic limit and isolating the optimal $\hat \Gamma$ operator.  That being said, for ascertaining the optimal form of $\hat \Gamma$, it will be essential to work with realistic {\em ab initio} potentials rather than models because, after all, there are physical limits to the differences in screening one can find in reality; finding a form for universal $\hat \Gamma$ that is accurate for all model Hamiltonians seems as unlikely as finding a DFT exchange-correlation functional \cite{hohenberg-kohn_DFT_1964} that is accurate for all model Hamiltonians. Clearly, if we wish to train such a potential, it will be crucial to work with realistic potentials. Luckily, we have already found quite a bit of success working with simple forms of $\hat \Gamma$ that can match VCD spectra\cite{tao_VCD_2024,duston_vcd_2024}, and our hope is that future progress will not be too difficult. 

Looking forward, the most exciting consequence of this manuscript is the possibility to use a PS framework and PS diabats to model electron transfer for problems that involve spin. Recent experiments have demonstrated chiral induced spin selectivity (CISS)\cite{naaman-waldeck_CISS-effect_2012,fransson_CISS_2020} for electron transfer, a phenomenon for which there is still no comprehensive fundamental theory. 
{As noted above, one strong advantage of single state PS over single state BH theory is that the former conserves total translational and angular momentum (allowing nuclei and electrons to exchange angular momentum) whereas the latter imposes the condition that the electronic translational and angular momentum is zero.} In practice, one can improve upon BO dynamics by including a Berry force\cite{berry_berry-curvature_1984,berry-robbins_berry-force_1993}, but such an approach is not really applicable for degenerate spin states, and our attempts to merge Berry forces with nonadiabatic dynamics semiclassically were not entirely successful (see Ref. \citenum{wu_FSSH-spin_2021} and compare with Ref. \citenum{wu_PSSH-spin_2022}). One might then hope that, if we model multiple states, we should find different spin couplings and branching ratios using a diabatic basis of PS states rather than BH states. Furthermore, obtaining accurate spin-dependent electron transfer rates in the presence of an external magnetic field is another important application\cite{bian_berry-force-rev_2023}. In short, the data presented here confirms that a PS approach can capture subtle differences in the spectrum of coupled nuclear-electronic systems by adding a new term to the electronic Hamiltonian that breaks time-reversal symmetry. And whereas this PS approach has recently been shown to recover $\Lambda-$doubling in small molecules for problems with nearly degenerate spin states\cite{peng_phase-space-electronic-structure_2025}, our hope is that the same approach can also provide new insight into CISS and spin-dependent ET.

\section{Data Availability}
The data that support the findings of this study are available from the corresponding author upon reasonable request. 

\section{Acknowledgments}
This work was supported by the U.S. Air Force Office of
Scientific Research (USAFOSR) under Grant No. FA9550-23-1-
0368.


\bibliography{mybib}

\end{document}